\documentclass[aps,prl,preprintnumbers,superscriptaddress,nofootinbib,tighten,twocolumn]{revtex4}
\usepackage{amssymb}
\usepackage{mathrsfs}
\usepackage{mathbbold}
\usepackage{graphicx}
\usepackage{amsmath}
\usepackage{epsfig}
\usepackage{mciteplus}
\usepackage{amsmath,color}
\usepackage{rotating,tabularx}

\newcommand\T{\rule{0pt}{2.8ex}} 
\newcommand\B{\rule[-1.2ex]{0pt}{0pt}} 
\newcommand{\ul}{\underline}

\def\slc#1{\setbox0=\hbox{$#1$}           
    \dimen0=\wd0                                 
    \setbox1=\hbox{/} \dimen1=\wd1               
    \ifdim\dimen0>\dimen1                        
       \rlap{\hbox to \dimen0{\hfil/\hfil}}      
       #1                                        
    \else                                        
       \rlap{\hbox to \dimen1{\hfil$#1$\hfil}}   
       /                                         
    \fi}
\begin{document}
\title{Light Sterile Neutrino in the Minimal Extended Seesaw}

\author{He Zhang}
\email{he.zhang@mpi-hd.mpg.de}

\affiliation{Max-Planck-Institut f{\"u}r Kernphysik, Saupfercheckweg
1, 69117 Heidelberg, Germany}


\begin{abstract}
\noindent Motivated by the recent observations on sterile neutrinos,
we present a minimal extension of the canonical type-I seesaw by
adding one extra singlet fermion. After the decoupling of
right-handed neutrinos, an eV-scale mass eigenstate is obtained
without the need of artificially inserting tiny mass scales or
Yukawa couplings for sterile neutrinos. In particular, the
active-sterile mixing is predicted to be of the order of 0.1.
Moreover, we show a concrete flavor $A_4$ model, in which the
required structures of the minimal extended seesaw are realized. We
also comment on the feasibility of accommodating a keV sterile
neutrino as an attractive candidate for warm dark matter.
\end{abstract}
\maketitle

\section{Introduction}

During the past decade, various neutrino oscillation experiments
have shown very solid evidence of non-vanishing neutrino masses and
lepton flavor mixing. Apart from neutrino oscillations within three
active flavors, recent re-evaluations of the anti-neutrino spectra
suggest that there exists a flux deficit in nuclear reactors, which
could be explained if anti-electron neutrinos oscillate to sterile
neutrinos~\cite{Mention:2011rk,*Huber:2011wv}. Such a picture would
require one or more sterile states with masses at the eV scale,
together with sizable admixtures [i.e., ${\cal O}(0.1)$] with active
neutrinos. Moreover, the light-element abundances from precision
cosmology and Big Bang nucleosynthesis favor extra radiation in the
Universe, which could be interpreted with the help of additional
sterile
neutrinos~\cite{Hamann:2010bk,*Izotov:2010ca,*Aver:2010wq,*Hamann:2011ge}.

From the theoretical side, it is unclear why the energy scales
related to electroweak symmetry breaking and sterile neutrinos are
different by many order of magnitude. In the canonical type-I
seesaw~\cite{Minkowski:1977sc,*Yanagida:1979as,*GellMann:1980vs,*Mohapatra:1979ia,*Schechter:1980gr},
right-handed neutrinos could in principle play the role of sterile
neutrinos if their masses lie in the eV ranges. This could be
technically natural since right-handed neutrinos are Standard Model
(SM) gauge singlets~\cite{deGouvea:2005er}. However, in such a case,
the Yukawa couplings relating lepton doublets and right-handed
neutrinos should be of the order $10^{-12}$ (namely, the Dirac mass
should be at the sub-eV scale) for correct mixings between active
and sterile neutrinos. It is therefore more appealing to consider a
natural and consistent framework yielding both low-scale sterile
neutrino masses and sizable active-sterile mixings. In this respect,
models simultaneously suppressing the Majorana and Dirac mass terms
have been proposed in the literature, e.g. the split seesaw models
in extra dimensions~\cite{Kusenko:2010ik,*Adulpravitchai:2011rq},
the Froggatt-Nielsen
mechanism~\cite{Froggatt:1978nt,Barry:2011wb,Merle:2011yv,Barry:2011fp},
and flavor
symmetries~\cite{Mohapatra:2001ns,*Shaposhnikov:2006nn,*Lindner:2010wr}.\footnote{See
also discussions in Ref.~\cite{Babu:2003is,*Babu:2004mj}.}

Recall that the seesaw mechanism is among one of the most popular
theoretical attempts that gives a natural way to understand the
smallness of neutrino masses. This motivates us to look for the
possibility of generating eV-scale sterile neutrino masses by using
a similar approach. Such an idea has been briefly mentioned in
Ref.~\cite{Barry:2011wb}, in which the type-I seesaw is extended by
adding only one singlet fermion [i.e., the minimal extended seesaw
(MES)] acting as a sterile neutrino, without the need of imposing
tiny Yukawa couplings or mass scales. A similar idea was also
employed in Ref.~\cite{Chun:1995js} to accommodate a sterile
neutrino of mass $\sim 10^{-3}~{\rm eV}$ in order to explain the
solar neutrino problem. In this note, we exploit in detail the
properties of the MES. Especially, we will show that the sterile
neutrino mass is stabilized at the eV scale, while a sizable
active-sterile mixing accounting for the rector neutrino anomaly is
predicted. Furthermore, we will discuss how to realize the MES
structure in flavor symmetries, i.e., a flavor model based on the
tetrahedral group $A_4$. We also comment that the model could be
generalized in order to include a keV sterile neutrino playing the
role of warm dark matter.

\section{The minimal extended seesaw}\label{sect:2}

Here we describe the basic structure of the MES, in which three
right-handed neutrinos and one additional gauge singlet chiral field
$S$ are introduced besides the SM particle content. We will show
that there could be a natural eV-scale sterile neutrino in this
picture, without the need of inserting a small mass term or tiny
Yukawa couplings. Explicitly, the Lagrangian of neutrino mass terms
is given by
\begin{eqnarray}\label{eq:L5}
-{\cal L}_m = \overline{\nu_L} M_D \nu_R  + \overline{S^c} M_S \nu_R
+ \frac{1}{2}\overline{\nu^c_R} M_R \nu_R +  {\rm h.c.}  ,
\end{eqnarray}
where $M_D$ and $M_R$ denote the Dirac and Majorana mass matrices,
respectively. Note that, $M_S$ is a $1 \times 3 $ matrix, since we
only introduce one extra singlet. The full $7 \times 7$ neutrino
mass matrix in the basis $(\nu_L,\nu^c_R,S^c)$ reads
\begin{eqnarray}
M_\nu^{7 \times 7}  = \left(\begin{matrix}  0 & M_D & 0 \cr M^T_D &
M_R & M^T_S \cr 0 & M_S & 0
\end{matrix} \right)  .
\end{eqnarray}
Similar to the typical type-I seesaw model, $M_D$ is assumed to be
around the electroweak scale, i.e., $10^2~{\rm GeV}$, while the
right-handed neutrino Majorana masses are chosen to be not far away
from the typical grand unification scale, $M_R \sim 10^{14}~{\rm
GeV}$. Furthermore, there is no bare Majorana mass term assumed for
$S$, while $S$ is only involved in the $M_S$ term, which may
originate from certain Yukawa interactions with right-handed
neutrinos and a SM singlet scalar. There is essentially no
constraint on the scale of $M_S$. In the remaining parts, we will
consider the interesting situation $M_R \gg M_S$.

In analogy to the type-I seesaw, the right-handed neutrinos are much
heavier than the electroweak scale, and thus they should be
decoupled at low scales. Effectively, one can block-diagonalize the
full mass matrix $M^{7\times 7}_\nu$ by using the seesaw formula,
and arrive at a $4\times 4$ neutrino mass matrix in the basis
$(\nu_L,S^c)$, i.e.,
\begin{eqnarray}\label{eq:M4times4}
M_\nu^{4\times 4} = - \left(\begin{matrix}  M_D  M^{-1}_R  M^T_D &
M_D  M^{-1}_R  M^T_S  \cr M_S  \left(M^{-1}_R\right)^T  M^T_D & M_S
M^{-1}_R M^T_S
\end{matrix} \right) .
\end{eqnarray}
One observes from Eq.~\eqref{eq:M4times4} that there exist in total
four light eigenstates corresponding to three active neutrinos and
one sterile neutrino, and their masses are all suppressed by a
factor $M^{-1}_R$ in the spirit of the seesaw mechanism. Moreover,
$M_\nu^{4\times4}$ is at most of rank 3, since {\footnotesize
\begin{eqnarray}\label{eq:det}
{\rm det}\left(M_\nu^{4\times4}\right)&=&{\rm det}\left(M_D M^{-1}_R
M^T_D  \right) {\rm det} \left[-M_S M^{-1}_R M^T_S \right. \nonumber
\\ &&+\left.  M_S M^{-1}_R
M^T_D \left(M_D  M^{-1}_R  M^T_D\right)^{-1} M_D  M^{-1}_R M^T_S
\right] \nonumber \\
&=& {\rm det}\left(M_D M^{-1}_R M^T_D  \right) {\rm det} \left[ M_S \left( M^{-1}_R - M^{-1}_R \right)M^T_S \right] \nonumber \\
&=& 0 \; ,
\end{eqnarray}}where we have assumed both $M_R$ and $M_D$ are invertible.
Therefore, at least one of the light neutrinos is massless.

We proceed to diagonalize $M_\nu^{4\times 4}$. There could be in
general three choices of the scale of $M_S$: 1) $M_D \sim M_S$; 2)
$M_D > M_S$; 3) $M_D < M_S$. For case 1, $M_\nu^{4\times 4}$ is
nearly democratic, indicating a maximal mixing between active and
sterile neutrinos, and therefore is not compatible with neutrino
oscillation data. In the second case, active neutrinos are heavier
than the sterile one. Such a scenario results in more tension with
cosmological constraints on the summation of light neutrino masses.
We will comment on this case later on. In what follows, we shall
concentrate on the third case, and study the properties of the
sterile neutrino in detail.

Since in case 3 $M_S$ is larger than $M_D$ by definition, one can
apply the seesaw formula once again to Eq.~\eqref{eq:M4times4}, and
obtain at leading order the active neutrino mass matrix
\begin{eqnarray}\label{eq:mnu}
m_\nu &\simeq& M_D  M^{-1}_R  M^T_S \left(M_S  M^{-1}_R
M^T_S\right)^{-1} M_S  \left(M^{-1}_R\right)^T  M^T_D \nonumber \\
& - & M_D M^{-1}_R M^T_D \, ,
\end{eqnarray}
as well as the sterile neutrino mass
\begin{eqnarray}\label{eq:ms}
m_s \simeq - M_S  M^{-1}_R M^T_S  \, .
\end{eqnarray}
Note that the right-hand-side of Eq.~\eqref{eq:mnu} does not vanish
since $M_S$ is a vector rather than a square matrix; if $M_S$ were a
square matrix this would lead to an exact cancellation between the
two terms of Eq.~\eqref{eq:mnu}. Here, $m_\nu$ can be diagonalized
by means of a $3\times 3$ unitary matrix as
\begin{eqnarray}\label{eq:ms}
m_\nu = U {\rm diag} (m_1,m_2,m_3) U^T \; ,
\end{eqnarray}
where $m_i$ (for $i=1,2,3$) denote the masses of three active
neutrinos. The full neutrino mixing matrix then takes a $4\times 4$
form~\cite{Schechter:1981cv},
\begin{eqnarray}
V \simeq  \left(\begin{matrix} (1-\frac{1}{2}RR^\dagger)U & R \cr
-R^\dagger U & 1 -\frac{1}{2} R^\dagger R
\end{matrix} \right)   ,
\end{eqnarray}
where $R$ is a $3\times 1$ matrix governing the strength of
active-sterile mixing,
\begin{eqnarray}\label{eq:R}
R= M_D  M^{-1}_R  M^T_S \left( M_S  M^{-1}_R M^T_S\right)^{-1} \; .
\end{eqnarray}
Essentially, the $R$ matrix (i.e., $V_{\alpha 4}$ for $\alpha
=e,\mu,\tau$) is suppressed by the ratio ${\cal O}(M_D)/{\cal
O}(M_S)$.

As a naive numerical example, for $M_D \sim 10^2~{\rm GeV}$, $M_S
\sim 5\times 10^2~{\rm GeV}$ and \mbox{$M_R \sim 2\times
10^{14}~{\rm GeV}$}, one may estimate that $m_\nu \sim 0.05~{\rm
eV}$, $m_s\sim 1.3~{\rm eV}$ together with $R \sim 0.2$. This is in
very good agreement with the fitted sterile neutrino
parameters~\cite{Kopp:2011qd,Giunti:2011hn}. We further stress that
the MES structure is stable against radiative corrections since the
loop contributions to the light neutrino masses are suppressed by
both the heavy right-handed neutrino masses and loop factors.

The MES picture described above is a minimal extension of the type-I
seesaw in the sense that one could allow for at most one extra
singlet in order to account for neutrino oscillation phenomena (see
a recent analysis in Ref.~\cite{Donini:2011jh}). In other words,
three heavy right-handed neutrinos can lead to at most three massive
light neutrinos (the ``seesaw-fair-play-rule''~\cite{Xing:2007uq}),
out of which two are active and needed to account for the solar and
atmospheric neutrino mixing.

Unfortunately, it is not possible to accommodate two eV-scale
sterile neutrinos in the MES, unless the number of right-handed
neutrinos is increased. Furthermore, if this scenario is embedded
into certain grand unified theory framework, e.g. $SO(10)$, it
cannot be anomaly free due to the lacking of other two generations
of $S$. Apart from these shortcomings, the MES possesses the
following features:

\begin{itemize}

\item apart from the electroweak and seesaw scales, one does not
artificially insert small mass scales for sterile neutrino masses.
As in the canonical type-I seesaw, one can take $M_S > M_D \sim
{\cal O}(10^2 ~{\rm GeV})$, while $M_R$ can be chosen close to the
$B-L$ scale, not far from the grand unification scale;

\item it is more predictive owing to the absence of one active neutrino
mass, while it does explain all the experimental data. Neutrino-less
double beta decay is also allowed because not all of the neutrinos
are light;

\item there exist heavy right-handed neutrinos that could be
responsible for thermal leptogenesis. Note that, in the setup we
considered, right-handed neutrinos would preferably decay to the
sterile neutrino since their coupling to $S$ is larger than active
neutrinos. However, this drawback could easily be circumvented since
$S$ enters in the one-loop self-energy diagram of the decay of
right-handed neutrinos, which could compensate for this.

\end{itemize}

We finally comment on the second case, i.e., $M_D \gg M_S$. Now that
$M^{4\times 4}_\nu$ possesses a hierarchical texture along the {\it
inverted} direction, one may still apply the seesaw formula to
Eq.~\eqref{eq:M4times4}, and obtain that, at leading order, the
active neutrino mass matrix is the same as that given in the type-I
seesaw, i.e., $m_\nu \simeq -M_D M^{-1}_R M^T_D$, whereas the
sterile neutrino mass is vanishing. In viewing of the experimental
results on the active-sterile mass-squared difference, one would
expect all the three active neutrinos to be located at the eV scale,
which is however challenged by standard cosmology since that leads
to a large total mass of neutrinos. Note also that, despite the
cosmological constraints, one can in principle add more singlets
since they do not affect active neutrino masses. Especially, in case
of three additional singlets $(S_1,S_2,S_3)$, the particle contents
are analogous to those in the inverse seesaw or double
seesaw~\cite{Mohapatra:1986aw,*Mohapatra:1986bd}, although the mass
matrix structures are clearly different.

\section{Realization in a flavor $A_4$ model}

In this section, we focus on a simple flavor $A_4$ model giving rise
to the exact mass structures depicted in Eq.~\eqref{eq:L5}. In
addition to the SM Higgs boson, we introduce three sets of flavons
$\varphi$, $\xi$ and $\chi$. An extra discrete abelian symmetry
$Z_4$ has been introduced in order to avoid interferences between
the neutrino and charged-lepton sectors. The particle assignments
are shown in Table~\ref{table:table-I}.
\begin{table}[h] \vspace{-0.15cm}
\centering \caption{Particle assignments in the flavor $A_4$ model.}
\label{table:table-I}  \vspace{0.2cm}
\begin{tabular}{c|ccccc|cccccc|cccc}
\hline \hline \T \B Field & $\ell$ & $e_R$ & $\mu_R$ & $\tau_R$ & $H$ & $\varphi$ & $\varphi'$ & $\varphi''$ & $\xi$ & $\xi'$ & $\chi$ & $\nu_{R1}$ & $\nu_{R2}$ & $\nu_{R3}$ & $S$ \\
\hline \T SU(2) & $2$ & $1$ & $1$ & $1$ & $2$  & $1$ & $1$ & $1$ & $1$ & $1$ & $1$ & $1$ & $1$ & $1$ & $1$\\
$A_4$ & $\ul{3}$ & $\ul{1}$ & $\ul{1}''$ & $\ul{1}'$ & $\ul{1}$ & $\ul{3}$ & $\ul{3}$ & $\ul{3}$ & $\ul{1}$ & $\ul{1}'$ & $\ul{1}$ & $\ul{1}$ & $\ul{1}'$ & $\ul{1}$ & $\ul{1}$ \\
$Z_4$ & $1$ & $1$ & $1$ & $1$ & $1$ & $1$ & ${\rm i}$ & $-1$ & $1$ & $-1$ & $-{\rm i}$ & $1$ & $-{\rm i}$ & $-1$ & $\rm i$ \\
\hline \hline
\end{tabular}\vspace{0.1cm}
\end{table}

At leading order, the $A_4 \otimes Z_4$ invariant Lagrangian for the
lepton sector is given by\footnote{Here the non-renormalizable
interactions for the neutrino sector are not included since they are
suppressed by $1/\Lambda$. In principle, dimension-five operators
like $\frac{1}{\Lambda}\varphi'\varphi' SS$ may spoil the desired
MES structure, since it leads to an unacceptable large mass term for
$S$ after the symmetry breaking. Such a drawback could be easily
avoided by introducing an additional global $U(1)_F$ symmetry, under
which only $\chi$ and $S$ are charged but with opposite sign. The
$\varphi'\varphi' SS$ term is then forbidden by $U(1)_F$, whereas
the $\chi S \nu_R$ interaction in the Lagrangian remains.}{\small
\begin{eqnarray}
{\cal L} & = &
\frac{y_e}{\Lambda}\left(\overline{\ell}H\varphi\right)_{\ul{1}} e_R
+
\frac{y_\mu}{\Lambda}\left(\overline{\ell}H\varphi\right)_{\ul{1}'}
\mu_R +
\frac{y_\tau}{\Lambda}\left(\overline{\ell}H\varphi\right)_{\ul{1}''}
\tau_R  \nonumber \\ &+&
\frac{y_1}{\Lambda}\left(\overline{\ell}\tilde
H\varphi\right)_{\ul{1}} \nu_{R1}+
\frac{y_2}{\Lambda}\left(\overline{\ell}\tilde
H\varphi'\right)_{\ul{1}''} \nu_{R2}+
\frac{y_3}{\Lambda}\left(\overline{\ell}\tilde
H\varphi''\right)_{\ul{1}} \nu_{R3} \nonumber \\ & + & \frac{1}{2}
\lambda_1 \xi \overline{\nu^c_{R1}}\nu_{R1} + \frac{1}{2} \lambda_2
\xi' \overline{\nu^c_{R2}}\nu_{R2} + \frac{1}{2} \lambda_3 \xi
\overline{\nu^c_{R3}}\nu_{R3} \nonumber \\ &+& \frac{1}{2} \rho \chi
\overline{S^c}\nu_{R1} + {\rm h.c.} \; ,
\end{eqnarray}}where $\Lambda$ denotes the cut-off scale and $\tilde H \equiv {\rm
i}\tau_2 H$. If we choose the real basis for $A_4$, along with the
flavon alignments\footnote{We do not expand our discussions on how
the vacuum alignment of flavons is achieved, whereas we refer
readers to Ref.~\cite{King:2006np}, in which the same flavon vacuum
alignment is acquired by assuming a radiative symmetry breaking
mechanism.}
\begin{eqnarray}
&&\langle \varphi \rangle =(v,0,0) \; , \quad \langle \varphi'
\rangle = (v,v,v) \; , \quad
\langle \varphi'' \rangle = (0,-v,v) \; , \nonumber \\
&&\langle \xi \rangle = \langle \xi' \rangle = v \; , \quad  \langle
\chi \rangle = u \; ,
\end{eqnarray}
then the charged-lepton mass matrix is diagonal\footnote{Note that,
the hierarchies between charged-lepton masses can be obtained by
using the Froggatt--Nielsen mechanism, viz., assigning different
$U(1)_{\rm FN}$ charges to the right-handed fields.}, i.e.,
\begin{eqnarray}
m_\ell =\frac{\langle H \rangle v}{\Lambda}{\rm diag}\left( y_e ,
y_\mu , y_\tau\right) \; ,
\end{eqnarray}
while the Dirac mass term is given by
\begin{eqnarray}
M_D = \frac{\langle H \rangle v}{\Lambda} \begin{pmatrix} y_1 & y_2
& 0 \cr 0 & y_2 & y_3 \cr 0 & y_2 & -y_3 \end{pmatrix} \; .
\end{eqnarray}
Due to the additional $Z_4$ symmetry, the right-handed neutrino mass
matrix is diagonal as well, viz.
\begin{eqnarray}
M_R = {\rm diag} \left( \lambda_1  v, \lambda_2 v, \lambda_3 v
\right) \; .
\end{eqnarray}
Furthermore, the singlet fermion $S$ does not acquire a Majorana
mass term, at least at leading order. The coupling matrix between
$S$ and right-handed neutrinos reads
\begin{eqnarray}
M_S = \begin{pmatrix} \rho u & 0 & 0
\end{pmatrix} \; .
\end{eqnarray}

As a rough numerical example, we assume the following mass scales:
$v\simeq 10^{13}~{\rm GeV}$, $\Lambda \simeq 10^{14}~{\rm GeV}$, and
$u\simeq 10^2~{\rm GeV}$. One can then estimate that, by assuming
order 1 Yukawa couplings, the condition $M_R \gg M_S > M_D$ can be
satisfied. Compared to Eq.~\eqref{eq:mnu}, we obtain
\begin{eqnarray}
m_\nu = -\frac{\langle H \rangle^2 v}{\Lambda^2}\begin{pmatrix}
\frac{y^2_2}{\lambda_2} & \frac{y^2_2}{\lambda_2} &
\frac{y^2_2}{\lambda_2} \cr \frac{y^2_2}{\lambda_2} & \frac{y^2_2
\lambda_3 + y^2_3 \lambda_2}{\lambda_2\lambda_3} & \frac{y^2_2
\lambda_3-y^2_3 \lambda_2}{\lambda_2\lambda_3} \cr
\frac{y^2_2}{\lambda_2} & \frac{y^2_2 \lambda_3-y^2_3
\lambda_2}{\lambda_2\lambda_3} & \frac{y^2_2 \lambda_3 + y^2_3
\lambda_2}{\lambda_2\lambda_3}
\end{pmatrix} \; .
\end{eqnarray}
It is straightforward to see that $m_\nu$ features a $\mu-\tau$
symmetry, which generally predicts a maximal mixing in the $2-3$
sector and a vanishing $\theta_{13}$. Indeed, $m_\nu$ can be
analytically diagonalized by using the tri-bimaximal
mixing~\cite{Harrison:2002er,*Xing:2002sw,*Harrison:2002kp,*He:2003rm}
matrix $V_{\rm TB}$ as
\begin{eqnarray}
m_\nu =- V_{\rm TB} ~{\rm diag} \begin{pmatrix}0,\frac{3 y^2_2
\langle H \rangle^2 v}{\lambda_2 \Lambda^2},\frac{2 y^2_3 \langle H
\rangle^2 v}{\lambda_3 \Lambda^2}\end{pmatrix}~V^T_{\rm TB}\; ,
\end{eqnarray}
with $V_{\rm TB}$ being
\begin{eqnarray}
V_{\rm TB} = \begin{pmatrix} \frac{2}{\sqrt{6}} & \frac{1}{\sqrt{3}}
& 0 \cr -\frac{1}{\sqrt{6}} & \frac{1}{\sqrt{3}} &
\frac{1}{\sqrt{2}} \cr -\frac{1}{\sqrt{6}} & \frac{1}{\sqrt{3}} & -
\frac{1}{\sqrt{2}}\end{pmatrix} \; .
\end{eqnarray}
Therefore, the normal mass ordering ($m_1\ll m_2\ll m_3$) together
with the tri-bimaximal mixing pattern are obtained. Taking for
example $y_3 = 0.91$, $y_2 = 0.31$, and $\lambda_2=\lambda_3=1$, one
obtains $\Delta m^2_{21} \simeq 7.6\times 10^{-5}~{\rm eV}^2$ and
$\Delta m^2_{31} \simeq 2.5\times 10^{-3}~{\rm eV}^2$, being
consistent with current global-fit data of neutrino mass-squared
differences~\cite{Schwetz:2011zk}.

The sterile neutrino mass is obtained from Eq.~\eqref{eq:ms} as
\begin{eqnarray}
m_s \simeq \frac{\rho^2 u^2}{\lambda_1 v} \; .
\end{eqnarray}
Fitting to the sterile neutrino mass from a recent best-fit given
in~\cite{Kopp:2011qd}, one can get $m_s\simeq 1.2~{\rm eV}$
(corresponding to $\Delta m^2_{41}\simeq 1.5~{\rm eV}^2$) for
$\rho=1.1$ and $\lambda_1=1$. By choosing $y_1 =1$ and inserting the
above parameters to Eq.~\eqref{eq:R}, we arrive at the
active-sterile mixing, i.e.,
\begin{eqnarray}
R \simeq \begin{pmatrix} \frac{y_1  \langle H \rangle v}{\rho u
\Lambda} & 0 & 0\end{pmatrix}^T \simeq \begin{pmatrix} 0.16 & 0 &
0\end{pmatrix}^T \; ,
\end{eqnarray}
corresponding to $|V_{e4}|^2 \simeq 0.025$ together with $|V_{\mu
4}|=|V_{\tau 4}|=0$, in good agreement with the best-fit value of
active-sterile mixing~\cite{Schwetz:2011qt} in the four neutrino
mixing scenario. Therefore, in this simple model, both the
tri-bimaximal mixing pattern in the active neutrino sector and a
sizable active-sterile neutrino mixing are predicted without the
need of fine-tuning the Yukawa couplings.

Alternatively, the flavor model described above can be slightly
changed in order to admit the inverted mass ordering of active
neutrinos (i.e., $m_2 \gtrsim m_1 \gg m_3$). For this purpose, one
could instead take the VEV alignment $\langle \varphi'\rangle =
(2v,-v,-v)$, which retains the tri-bimaximal mixing in the active
neutrino mixing, and leads to a vanishing mass $m_3=0$. Note that,
in case of the inverted mass ordering, the next-to-leading seesaw
corrections~\cite{Hettmansperger:2011bt} should be included in the
diagonalization of $M^{4\times4}_\nu$, because of the degeneracy
between $m_1$ and $m_2$.

As mentioned in the previous section, in this model, both active and
sterile neutrinos may mediate the neutrino-less double beta decay
processes, and their contributions to the effective mass are not
cancelled. Concretely, we have $\langle m\rangle_{ee} \simeq |m_2
V^2_{e2}+m_s V_{e4}^2| \simeq m_s |V_{e4}|^2$ in the normal mass
ordering case, and $\langle m\rangle_{ee} \simeq |m_1 V^2_{e1}+m_2
V^2_{e2}+m_s V_{e4}^2| \simeq |m_1 + m_s V_{e4}^2 |$ in the inverted
mass ordering case. In addition, effects from right-handed neutrinos
are negligibly small since they are highly suppressed by $M_R$ (see
e.g. Ref.~\cite{Rodejohann:2011mu} for detailed discussions). This
is a very distinctive feature, in particular compared to models with
only eV-scale right-handed neutrinos, in which neutrino-less double
beta decays are forbidden.

One may also wonder if the model could be modified to allow for a
keV sterile neutrino warm dark matter candidate. Indeed, the sterile
neutrino mass can be chosen at the keV ranges by setting, e.g. $u
\sim 4 ~{\rm TeV}$. Using the same Yukawa coupling parameters in the
previous discussions, one then arrives at $m_s \simeq 1.9~{\rm
keV}$. Unfortunately, the active-sterile mixing $\theta_s = R_{11}
\simeq 4\times 10^{-3}$ turns out to exceed the current X-ray
constraint~\cite{Boyarsky:2009ix},
\begin{eqnarray}\label{eq:X-ray}
\theta^2_s \lesssim 1.8\times 10^{-5} \left(\frac{1~{\rm
keV}}{m_s}\right)^5 \; .
\end{eqnarray}
In order to keep $\theta_s$ small enough, we need a mild tuning of
the Yukawa coupling, i.e., $y_1 < 0.2$. For example, taking $y_1 =
0.15$, we get $\theta^2_s \simeq 1.2\times 10^{-5}\left(\frac{1~{\rm
keV}}{m_s}\right)^5$, satisfying the bound in
Eq.~\eqref{eq:X-ray}.\footnote{Note that various mechanisms for the
production of sterile neutrino warm dark matter have been proposed
in the literature, which, however, will not be discussed in this
note.}

Finally, we note that the reactor experiments Double
Chooz~\cite{Abe:2011fz}, Daya Bay~\cite{An:2012eh} and
RENO~\cite{Ahn:2012nd} has found the smallest mixing angle
$\theta_{13}$, i.e., $\sin^2\theta_{13} \simeq 0.025$ from a recent
global-fit~\cite{Fogli:2012ua}. Therefore, the exact tri-bimaximal
mixing pattern should be modified in order to accommodate
non-vanishing $\theta_{13}$. This can be achieved by including the
next-to-leading order corrections to the flavor model. For example,
in Ref.~\cite{Altarelli:2005yp}, the higher dimensional corrections
to the vacuum alignments would result in a sizable $\theta_{13}$
compatible with the current experimental observation. Moreover,
perturbations to the charged-lepton sector may also lead to a large
$\theta_{13}$ (see discussions in Ref.~\cite{Barry:2011fp}). Since
the main purpose of this work is to present a novel mechanism
generating light sterile neutrinos, we will not further expand our
discussion on $\theta_{13}$.

\section{Conclusion}

In this note, we have studied a minimal extension of the type-I
seesaw, which contains an extra singlet fermion coupled purely to
the right-handed neutrinos. In such a framework, both active and
sterile neutrino masses are suppressed via the seesaw mechanism, and
thus, an eV-scale sterile neutrino together with sizable
active-sterile mixing is accommodated without the need of
artificially inserting small mass scales or Yukawa couplings.
Furthermore, we have presented a flavor $A_4$ model, in which both
the MES structures and the tri-bimaximal mixing pattern are
realized. In particular, for a sterile neutrino with mass being
around the eV scale, the active-sterile mixing (i.e., $|V_{e4}|$) is
found to be of the order of $0.1$, in good agreement with current
experimental observations. The model may also be modified to take
keV sterile neutrino warm dark matter into account. We hope this
note serves as a useful guide for future model building works on
low-scale sterile neutrinos.

\

\begin{acknowledgments}
We would like to thank Werner Rodejohann, James Barry and Alexander
Merle for helpful discussions and comments on the manuscript. This
work was supported by the ERC under the Starting Grant MANITOP and
by the Deutsche Forschungsgemeinschaft in the Transregio 27
``Neutrinos and beyond -- weakly interacting particles in physics,
astrophysics and cosmology''.
\end{acknowledgments}

\bibliography{bib}

\ifx\mcitethebibliography\mciteundefinedmacro
\PackageError{apsrevM.bst}{mciteplus.sty has not been loaded}
{This bibstyle requires the use of the mciteplus package.}\fi
\begin{mcitethebibliography}{45}
\expandafter\ifx\csname natexlab\endcsname\relax\def\natexlab#1{#1}\fi
\expandafter\ifx\csname bibnamefont\endcsname\relax
  \def\bibnamefont#1{#1}\fi
\expandafter\ifx\csname bibfnamefont\endcsname\relax
  \def\bibfnamefont#1{#1}\fi
\expandafter\ifx\csname citenamefont\endcsname\relax
  \def\citenamefont#1{#1}\fi
\expandafter\ifx\csname url\endcsname\relax
  \def\url#1{\texttt{#1}}\fi
\expandafter\ifx\csname urlprefix\endcsname\relax\def\urlprefix{URL }\fi
\providecommand{\bibinfo}[2]{#2}
\providecommand{\eprint}[2][]{\url{#2}}

\bibitem[{\citenamefont{Mention et~al.}(2011)\citenamefont{Mention, Fechner,
  Lasserre, Mueller, Lhuillier et~al.}}]{Mention:2011rk}
\bibinfo{author}{\bibfnamefont{G.}~\bibnamefont{Mention}},
  \bibinfo{author}{\bibfnamefont{M.}~\bibnamefont{Fechner}},
  \bibinfo{author}{\bibfnamefont{T.}~\bibnamefont{Lasserre}},
  \bibinfo{author}{\bibfnamefont{T.}~\bibnamefont{Mueller}},
  \bibinfo{author}{\bibfnamefont{D.}~\bibnamefont{Lhuillier}},
  \bibnamefont{et~al.}, \bibinfo{journal}{Phys.Rev.}
  \textbf{\bibinfo{volume}{D83}}, \bibinfo{pages}{073006}
  (\bibinfo{year}{2011}), \eprint{1101.2755}\relax
\mciteBstWouldAddEndPuncttrue
\mciteSetBstMidEndSepPunct{\mcitedefaultmidpunct}
{\mcitedefaultendpunct}{\mcitedefaultseppunct}\relax
\EndOfBibitem
\bibitem[{\citenamefont{Huber}(2011)}]{Huber:2011wv}
\bibinfo{author}{\bibfnamefont{P.}~\bibnamefont{Huber}},
  \bibinfo{journal}{Phys.Rev.} \textbf{\bibinfo{volume}{C84}},
  \bibinfo{pages}{024617} (\bibinfo{year}{2011}), \eprint{1106.0687}\relax
\mciteBstWouldAddEndPuncttrue
\mciteSetBstMidEndSepPunct{\mcitedefaultmidpunct}
{\mcitedefaultendpunct}{\mcitedefaultseppunct}\relax
\EndOfBibitem
\bibitem[{\citenamefont{Hamann et~al.}(2010)\citenamefont{Hamann, Hannestad,
  Raffelt, Tamborra, and Wong}}]{Hamann:2010bk}
\bibinfo{author}{\bibfnamefont{J.}~\bibnamefont{Hamann}},
  \bibinfo{author}{\bibfnamefont{S.}~\bibnamefont{Hannestad}},
  \bibinfo{author}{\bibfnamefont{G.~G.} \bibnamefont{Raffelt}},
  \bibinfo{author}{\bibfnamefont{I.}~\bibnamefont{Tamborra}}, \bibnamefont{and}
  \bibinfo{author}{\bibfnamefont{Y.~Y.} \bibnamefont{Wong}},
  \bibinfo{journal}{Phys.Rev.Lett.} \textbf{\bibinfo{volume}{105}},
  \bibinfo{pages}{181301} (\bibinfo{year}{2010}), \eprint{1006.5276}\relax
\mciteBstWouldAddEndPuncttrue
\mciteSetBstMidEndSepPunct{\mcitedefaultmidpunct}
{\mcitedefaultendpunct}{\mcitedefaultseppunct}\relax
\EndOfBibitem
\bibitem[{\citenamefont{Izotov and Thuan}(2010)}]{Izotov:2010ca}
\bibinfo{author}{\bibfnamefont{Y.}~\bibnamefont{Izotov}} \bibnamefont{and}
  \bibinfo{author}{\bibfnamefont{T.}~\bibnamefont{Thuan}},
  \bibinfo{journal}{Astrophys.J.} \textbf{\bibinfo{volume}{710}},
  \bibinfo{pages}{L67} (\bibinfo{year}{2010}), \eprint{1001.4440}\relax
\mciteBstWouldAddEndPuncttrue
\mciteSetBstMidEndSepPunct{\mcitedefaultmidpunct}
{\mcitedefaultendpunct}{\mcitedefaultseppunct}\relax
\EndOfBibitem
\bibitem[{\citenamefont{Aver et~al.}(2010)\citenamefont{Aver, Olive, and
  Skillman}}]{Aver:2010wq}
\bibinfo{author}{\bibfnamefont{E.}~\bibnamefont{Aver}},
  \bibinfo{author}{\bibfnamefont{K.~A.} \bibnamefont{Olive}}, \bibnamefont{and}
  \bibinfo{author}{\bibfnamefont{E.~D.} \bibnamefont{Skillman}},
  \bibinfo{journal}{JCAP} \textbf{\bibinfo{volume}{1005}}, \bibinfo{pages}{003}
  (\bibinfo{year}{2010}), \eprint{1001.5218}\relax
\mciteBstWouldAddEndPuncttrue
\mciteSetBstMidEndSepPunct{\mcitedefaultmidpunct}
{\mcitedefaultendpunct}{\mcitedefaultseppunct}\relax
\EndOfBibitem
\bibitem[{\citenamefont{Hamann et~al.}(2011)\citenamefont{Hamann, Hannestad,
  Raffelt, and Wong}}]{Hamann:2011ge}
\bibinfo{author}{\bibfnamefont{J.}~\bibnamefont{Hamann}},
  \bibinfo{author}{\bibfnamefont{S.}~\bibnamefont{Hannestad}},
  \bibinfo{author}{\bibfnamefont{G.~G.} \bibnamefont{Raffelt}},
  \bibnamefont{and} \bibinfo{author}{\bibfnamefont{Y.~Y.} \bibnamefont{Wong}}
  (\bibinfo{year}{2011}), \eprint{1108.4136}\relax
\mciteBstWouldAddEndPuncttrue
\mciteSetBstMidEndSepPunct{\mcitedefaultmidpunct}
{\mcitedefaultendpunct}{\mcitedefaultseppunct}\relax
\EndOfBibitem
\bibitem[{\citenamefont{Minkowski}(1977)}]{Minkowski:1977sc}
\bibinfo{author}{\bibfnamefont{P.}~\bibnamefont{Minkowski}},
  \bibinfo{journal}{Phys. Lett.} \textbf{\bibinfo{volume}{B67}},
  \bibinfo{pages}{421} (\bibinfo{year}{1977})\relax
\mciteBstWouldAddEndPuncttrue
\mciteSetBstMidEndSepPunct{\mcitedefaultmidpunct}
{\mcitedefaultendpunct}{\mcitedefaultseppunct}\relax
\EndOfBibitem
\bibitem[{\citenamefont{Yanagida}(1979)}]{Yanagida:1979as}
\bibinfo{author}{\bibfnamefont{T.}~\bibnamefont{Yanagida}}, in
  \emph{\bibinfo{booktitle}{Proc. Workshop on the Baryon Number of the Universe
  and Unified Theories}}, edited by
  \bibinfo{editor}{\bibfnamefont{O.}~\bibnamefont{Sawada}} \bibnamefont{and}
  \bibinfo{editor}{\bibfnamefont{A.}~\bibnamefont{Sugamoto}}
  (\bibinfo{year}{1979}), p.~\bibinfo{pages}{95}\relax
\mciteBstWouldAddEndPuncttrue
\mciteSetBstMidEndSepPunct{\mcitedefaultmidpunct}
{\mcitedefaultendpunct}{\mcitedefaultseppunct}\relax
\EndOfBibitem
\bibitem[{\citenamefont{Gell-Mann et~al.}(1979)\citenamefont{Gell-Mann, Ramond,
  and Slansky}}]{GellMann:1980vs}
\bibinfo{author}{\bibfnamefont{M.}~\bibnamefont{Gell-Mann}},
  \bibinfo{author}{\bibfnamefont{P.}~\bibnamefont{Ramond}}, \bibnamefont{and}
  \bibinfo{author}{\bibfnamefont{R.}~\bibnamefont{Slansky}}, in
  \emph{\bibinfo{booktitle}{Supergravity}}, edited by
  \bibinfo{editor}{\bibfnamefont{P.}~\bibnamefont{{van Nieuwenhuizen}}}
  \bibnamefont{and} \bibinfo{editor}{\bibfnamefont{D.}~\bibnamefont{Freedman}}
  (\bibinfo{year}{1979}), p. \bibinfo{pages}{315}\relax
\mciteBstWouldAddEndPuncttrue
\mciteSetBstMidEndSepPunct{\mcitedefaultmidpunct}
{\mcitedefaultendpunct}{\mcitedefaultseppunct}\relax
\EndOfBibitem
\bibitem[{\citenamefont{Mohapatra and Senjanovic}(1980)}]{Mohapatra:1979ia}
\bibinfo{author}{\bibfnamefont{R.~N.} \bibnamefont{Mohapatra}}
  \bibnamefont{and}
  \bibinfo{author}{\bibfnamefont{G.}~\bibnamefont{Senjanovic}},
  \bibinfo{journal}{Phys. Rev. Lett.} \textbf{\bibinfo{volume}{44}},
  \bibinfo{pages}{912} (\bibinfo{year}{1980})\relax
\mciteBstWouldAddEndPuncttrue
\mciteSetBstMidEndSepPunct{\mcitedefaultmidpunct}
{\mcitedefaultendpunct}{\mcitedefaultseppunct}\relax
\EndOfBibitem
\bibitem[{\citenamefont{Schechter and Valle}(1980)}]{Schechter:1980gr}
\bibinfo{author}{\bibfnamefont{J.}~\bibnamefont{Schechter}} \bibnamefont{and}
  \bibinfo{author}{\bibfnamefont{J.}~\bibnamefont{Valle}},
  \bibinfo{journal}{Phys.Rev.} \textbf{\bibinfo{volume}{D22}},
  \bibinfo{pages}{2227} (\bibinfo{year}{1980})\relax
\mciteBstWouldAddEndPuncttrue
\mciteSetBstMidEndSepPunct{\mcitedefaultmidpunct}
{\mcitedefaultendpunct}{\mcitedefaultseppunct}\relax
\EndOfBibitem
\bibitem[{\citenamefont{de~Gouvea}(2005)}]{deGouvea:2005er}
\bibinfo{author}{\bibfnamefont{A.}~\bibnamefont{de~Gouvea}},
  \bibinfo{journal}{Phys.Rev.} \textbf{\bibinfo{volume}{D72}},
  \bibinfo{pages}{033005} (\bibinfo{year}{2005}), \eprint{hep-ph/0501039}\relax
\mciteBstWouldAddEndPuncttrue
\mciteSetBstMidEndSepPunct{\mcitedefaultmidpunct}
{\mcitedefaultendpunct}{\mcitedefaultseppunct}\relax
\EndOfBibitem
\bibitem[{\citenamefont{Kusenko et~al.}(2010)\citenamefont{Kusenko, Takahashi,
  and Yanagida}}]{Kusenko:2010ik}
\bibinfo{author}{\bibfnamefont{A.}~\bibnamefont{Kusenko}},
  \bibinfo{author}{\bibfnamefont{F.}~\bibnamefont{Takahashi}},
  \bibnamefont{and} \bibinfo{author}{\bibfnamefont{T.~T.}
  \bibnamefont{Yanagida}}, \bibinfo{journal}{Phys. Lett.}
  \textbf{\bibinfo{volume}{B693}}, \bibinfo{pages}{144} (\bibinfo{year}{2010}),
  \eprint{1006.1731}\relax
\mciteBstWouldAddEndPuncttrue
\mciteSetBstMidEndSepPunct{\mcitedefaultmidpunct}
{\mcitedefaultendpunct}{\mcitedefaultseppunct}\relax
\EndOfBibitem
\bibitem[{\citenamefont{Adulpravitchai and
  Takahashi}(2011)}]{Adulpravitchai:2011rq}
\bibinfo{author}{\bibfnamefont{A.}~\bibnamefont{Adulpravitchai}}
  \bibnamefont{and}
  \bibinfo{author}{\bibfnamefont{R.}~\bibnamefont{Takahashi}},
  \bibinfo{journal}{JHEP} \textbf{\bibinfo{volume}{09}}, \bibinfo{pages}{127}
  (\bibinfo{year}{2011}), \eprint{1107.3829}\relax
\mciteBstWouldAddEndPuncttrue
\mciteSetBstMidEndSepPunct{\mcitedefaultmidpunct}
{\mcitedefaultendpunct}{\mcitedefaultseppunct}\relax
\EndOfBibitem
\bibitem[{\citenamefont{Froggatt and Nielsen}(1979)}]{Froggatt:1978nt}
\bibinfo{author}{\bibfnamefont{C.}~\bibnamefont{Froggatt}} \bibnamefont{and}
  \bibinfo{author}{\bibfnamefont{H.~B.} \bibnamefont{Nielsen}},
  \bibinfo{journal}{Nucl.Phys.} \textbf{\bibinfo{volume}{B147}},
  \bibinfo{pages}{277} (\bibinfo{year}{1979})\relax
\mciteBstWouldAddEndPuncttrue
\mciteSetBstMidEndSepPunct{\mcitedefaultmidpunct}
{\mcitedefaultendpunct}{\mcitedefaultseppunct}\relax
\EndOfBibitem
\bibitem[{\citenamefont{Barry et~al.}(2011{\natexlab{a}})\citenamefont{Barry,
  Rodejohann, and Zhang}}]{Barry:2011wb}
\bibinfo{author}{\bibfnamefont{J.}~\bibnamefont{Barry}},
  \bibinfo{author}{\bibfnamefont{W.}~\bibnamefont{Rodejohann}},
  \bibnamefont{and} \bibinfo{author}{\bibfnamefont{H.}~\bibnamefont{Zhang}},
  \bibinfo{journal}{JHEP} \textbf{\bibinfo{volume}{1107}}, \bibinfo{pages}{091}
  (\bibinfo{year}{2011}{\natexlab{a}}), \eprint{1105.3911}\relax
\mciteBstWouldAddEndPuncttrue
\mciteSetBstMidEndSepPunct{\mcitedefaultmidpunct}
{\mcitedefaultendpunct}{\mcitedefaultseppunct}\relax
\EndOfBibitem
\bibitem[{\citenamefont{Merle and Niro}(2011)}]{Merle:2011yv}
\bibinfo{author}{\bibfnamefont{A.}~\bibnamefont{Merle}} \bibnamefont{and}
  \bibinfo{author}{\bibfnamefont{V.}~\bibnamefont{Niro}},
  \bibinfo{journal}{JCAP} \textbf{\bibinfo{volume}{1107}}, \bibinfo{pages}{023}
  (\bibinfo{year}{2011}), \eprint{1105.5136}\relax
\mciteBstWouldAddEndPuncttrue
\mciteSetBstMidEndSepPunct{\mcitedefaultmidpunct}
{\mcitedefaultendpunct}{\mcitedefaultseppunct}\relax
\EndOfBibitem
\bibitem[{\citenamefont{Barry et~al.}(2011{\natexlab{b}})\citenamefont{Barry,
  Rodejohann, and Zhang}}]{Barry:2011fp}
\bibinfo{author}{\bibfnamefont{J.}~\bibnamefont{Barry}},
  \bibinfo{author}{\bibfnamefont{W.}~\bibnamefont{Rodejohann}},
  \bibnamefont{and} \bibinfo{author}{\bibfnamefont{H.}~\bibnamefont{Zhang}}
  (\bibinfo{year}{2011}{\natexlab{b}}), \eprint{1110.6382}\relax
\mciteBstWouldAddEndPuncttrue
\mciteSetBstMidEndSepPunct{\mcitedefaultmidpunct}
{\mcitedefaultendpunct}{\mcitedefaultseppunct}\relax
\EndOfBibitem
\bibitem[{\citenamefont{Mohapatra}(2001)}]{Mohapatra:2001ns}
\bibinfo{author}{\bibfnamefont{R.}~\bibnamefont{Mohapatra}},
  \bibinfo{journal}{Phys.Rev.} \textbf{\bibinfo{volume}{D64}},
  \bibinfo{pages}{091301} (\bibinfo{year}{2001}), \eprint{hep-ph/0107264}\relax
\mciteBstWouldAddEndPuncttrue
\mciteSetBstMidEndSepPunct{\mcitedefaultmidpunct}
{\mcitedefaultendpunct}{\mcitedefaultseppunct}\relax
\EndOfBibitem
\bibitem[{\citenamefont{Shaposhnikov}(2007)}]{Shaposhnikov:2006nn}
\bibinfo{author}{\bibfnamefont{M.}~\bibnamefont{Shaposhnikov}},
  \bibinfo{journal}{Nucl.Phys.} \textbf{\bibinfo{volume}{B763}},
  \bibinfo{pages}{49} (\bibinfo{year}{2007}), \eprint{hep-ph/0605047}\relax
\mciteBstWouldAddEndPuncttrue
\mciteSetBstMidEndSepPunct{\mcitedefaultmidpunct}
{\mcitedefaultendpunct}{\mcitedefaultseppunct}\relax
\EndOfBibitem
\bibitem[{\citenamefont{Lindner et~al.}(2011)\citenamefont{Lindner, Merle, and
  Niro}}]{Lindner:2010wr}
\bibinfo{author}{\bibfnamefont{M.}~\bibnamefont{Lindner}},
  \bibinfo{author}{\bibfnamefont{A.}~\bibnamefont{Merle}}, \bibnamefont{and}
  \bibinfo{author}{\bibfnamefont{V.}~\bibnamefont{Niro}},
  \bibinfo{journal}{JCAP} \textbf{\bibinfo{volume}{1101}}, \bibinfo{pages}{034}
  (\bibinfo{year}{2011}), \eprint{1011.4950}\relax
\mciteBstWouldAddEndPuncttrue
\mciteSetBstMidEndSepPunct{\mcitedefaultmidpunct}
{\mcitedefaultendpunct}{\mcitedefaultseppunct}\relax
\EndOfBibitem
\bibitem[{\citenamefont{Babu and Seidl}(2004{\natexlab{a}})}]{Babu:2003is}
\bibinfo{author}{\bibfnamefont{K.~S.} \bibnamefont{Babu}} \bibnamefont{and}
  \bibinfo{author}{\bibfnamefont{G.}~\bibnamefont{Seidl}},
  \bibinfo{journal}{Phys. Lett.} \textbf{\bibinfo{volume}{B591}},
  \bibinfo{pages}{127} (\bibinfo{year}{2004}{\natexlab{a}}),
  \eprint{hep-ph/0312285}\relax
\mciteBstWouldAddEndPuncttrue
\mciteSetBstMidEndSepPunct{\mcitedefaultmidpunct}
{\mcitedefaultendpunct}{\mcitedefaultseppunct}\relax
\EndOfBibitem
\bibitem[{\citenamefont{Babu and Seidl}(2004{\natexlab{b}})}]{Babu:2004mj}
\bibinfo{author}{\bibfnamefont{K.~S.} \bibnamefont{Babu}} \bibnamefont{and}
  \bibinfo{author}{\bibfnamefont{G.}~\bibnamefont{Seidl}},
  \bibinfo{journal}{Phys. Rev.} \textbf{\bibinfo{volume}{D70}},
  \bibinfo{pages}{113014} (\bibinfo{year}{2004}{\natexlab{b}}),
  \eprint{hep-ph/0405197}\relax
\mciteBstWouldAddEndPuncttrue
\mciteSetBstMidEndSepPunct{\mcitedefaultmidpunct}
{\mcitedefaultendpunct}{\mcitedefaultseppunct}\relax
\EndOfBibitem
\bibitem[{\citenamefont{Chun et~al.}(1995)\citenamefont{Chun, Joshipura, and
  Smirnov}}]{Chun:1995js}
\bibinfo{author}{\bibfnamefont{E.}~\bibnamefont{Chun}},
  \bibinfo{author}{\bibfnamefont{A.~S.} \bibnamefont{Joshipura}},
  \bibnamefont{and} \bibinfo{author}{\bibfnamefont{A.}~\bibnamefont{Smirnov}},
  \bibinfo{journal}{Phys.Lett.} \textbf{\bibinfo{volume}{B357}},
  \bibinfo{pages}{608} (\bibinfo{year}{1995}), \eprint{hep-ph/9505275}\relax
\mciteBstWouldAddEndPuncttrue
\mciteSetBstMidEndSepPunct{\mcitedefaultmidpunct}
{\mcitedefaultendpunct}{\mcitedefaultseppunct}\relax
\EndOfBibitem
\bibitem[{\citenamefont{Schechter and Valle}(1982)}]{Schechter:1981cv}
\bibinfo{author}{\bibfnamefont{J.}~\bibnamefont{Schechter}} \bibnamefont{and}
  \bibinfo{author}{\bibfnamefont{J.}~\bibnamefont{Valle}},
  \bibinfo{journal}{Phys.Rev.} \textbf{\bibinfo{volume}{D25}},
  \bibinfo{pages}{774} (\bibinfo{year}{1982})\relax
\mciteBstWouldAddEndPuncttrue
\mciteSetBstMidEndSepPunct{\mcitedefaultmidpunct}
{\mcitedefaultendpunct}{\mcitedefaultseppunct}\relax
\EndOfBibitem
\bibitem[{\citenamefont{Kopp et~al.}(2011)\citenamefont{Kopp, Maltoni, and
  Schwetz}}]{Kopp:2011qd}
\bibinfo{author}{\bibfnamefont{J.}~\bibnamefont{Kopp}},
  \bibinfo{author}{\bibfnamefont{M.}~\bibnamefont{Maltoni}}, \bibnamefont{and}
  \bibinfo{author}{\bibfnamefont{T.}~\bibnamefont{Schwetz}},
  \bibinfo{journal}{Phys.Rev.Lett.} \textbf{\bibinfo{volume}{107}},
  \bibinfo{pages}{091801} (\bibinfo{year}{2011}), \eprint{1103.4570}\relax
\mciteBstWouldAddEndPuncttrue
\mciteSetBstMidEndSepPunct{\mcitedefaultmidpunct}
{\mcitedefaultendpunct}{\mcitedefaultseppunct}\relax
\EndOfBibitem
\bibitem[{\citenamefont{Giunti and Laveder}(2011)}]{Giunti:2011hn}
\bibinfo{author}{\bibfnamefont{C.}~\bibnamefont{Giunti}} \bibnamefont{and}
  \bibinfo{author}{\bibfnamefont{M.}~\bibnamefont{Laveder}}
  (\bibinfo{year}{2011}), \eprint{1109.4033}\relax
\mciteBstWouldAddEndPuncttrue
\mciteSetBstMidEndSepPunct{\mcitedefaultmidpunct}
{\mcitedefaultendpunct}{\mcitedefaultseppunct}\relax
\EndOfBibitem
\bibitem[{\citenamefont{Donini et~al.}(2011)\citenamefont{Donini, Hernandez,
  Lopez-Pavon, and Maltoni}}]{Donini:2011jh}
\bibinfo{author}{\bibfnamefont{A.}~\bibnamefont{Donini}},
  \bibinfo{author}{\bibfnamefont{P.}~\bibnamefont{Hernandez}},
  \bibinfo{author}{\bibfnamefont{J.}~\bibnamefont{Lopez-Pavon}},
  \bibnamefont{and} \bibinfo{author}{\bibfnamefont{M.}~\bibnamefont{Maltoni}},
  \bibinfo{journal}{JHEP} \textbf{\bibinfo{volume}{1107}}, \bibinfo{pages}{105}
  (\bibinfo{year}{2011}), \eprint{1106.0064}\relax
\mciteBstWouldAddEndPuncttrue
\mciteSetBstMidEndSepPunct{\mcitedefaultmidpunct}
{\mcitedefaultendpunct}{\mcitedefaultseppunct}\relax
\EndOfBibitem
\bibitem[{\citenamefont{Xing}(2008)}]{Xing:2007uq}
\bibinfo{author}{\bibfnamefont{Z.-Z.} \bibnamefont{Xing}},
  \bibinfo{journal}{Chin.Phys.} \textbf{\bibinfo{volume}{C32}},
  \bibinfo{pages}{96} (\bibinfo{year}{2008}), \eprint{0706.0052}\relax
\mciteBstWouldAddEndPuncttrue
\mciteSetBstMidEndSepPunct{\mcitedefaultmidpunct}
{\mcitedefaultendpunct}{\mcitedefaultseppunct}\relax
\EndOfBibitem
\bibitem[{\citenamefont{Mohapatra}(1986)}]{Mohapatra:1986aw}
\bibinfo{author}{\bibfnamefont{R.}~\bibnamefont{Mohapatra}},
  \bibinfo{journal}{Phys.Rev.Lett.} \textbf{\bibinfo{volume}{56}},
  \bibinfo{pages}{561} (\bibinfo{year}{1986})\relax
\mciteBstWouldAddEndPuncttrue
\mciteSetBstMidEndSepPunct{\mcitedefaultmidpunct}
{\mcitedefaultendpunct}{\mcitedefaultseppunct}\relax
\EndOfBibitem
\bibitem[{\citenamefont{Mohapatra and Valle}(1986)}]{Mohapatra:1986bd}
\bibinfo{author}{\bibfnamefont{R.~N.} \bibnamefont{Mohapatra}}
  \bibnamefont{and} \bibinfo{author}{\bibfnamefont{J.~W.~F.}
  \bibnamefont{Valle}}, \bibinfo{journal}{Phys. Rev.}
  \textbf{\bibinfo{volume}{D34}}, \bibinfo{pages}{1642}
  (\bibinfo{year}{1986})\relax
\mciteBstWouldAddEndPuncttrue
\mciteSetBstMidEndSepPunct{\mcitedefaultmidpunct}
{\mcitedefaultendpunct}{\mcitedefaultseppunct}\relax
\EndOfBibitem
\bibitem[{\citenamefont{King and Malinsky}(2007)}]{King:2006np}
\bibinfo{author}{\bibfnamefont{S.~F.} \bibnamefont{King}} \bibnamefont{and}
  \bibinfo{author}{\bibfnamefont{M.}~\bibnamefont{Malinsky}},
  \bibinfo{journal}{Phys. Lett.} \textbf{\bibinfo{volume}{B645}},
  \bibinfo{pages}{351} (\bibinfo{year}{2007}), \eprint{hep-ph/0610250}\relax
\mciteBstWouldAddEndPuncttrue
\mciteSetBstMidEndSepPunct{\mcitedefaultmidpunct}
{\mcitedefaultendpunct}{\mcitedefaultseppunct}\relax
\EndOfBibitem
\bibitem[{\citenamefont{Harrison et~al.}(2002)\citenamefont{Harrison, Perkins,
  and Scott}}]{Harrison:2002er}
\bibinfo{author}{\bibfnamefont{P.~F.} \bibnamefont{Harrison}},
  \bibinfo{author}{\bibfnamefont{D.~H.} \bibnamefont{Perkins}},
  \bibnamefont{and} \bibinfo{author}{\bibfnamefont{W.~G.} \bibnamefont{Scott}},
  \bibinfo{journal}{Phys. Lett.} \textbf{\bibinfo{volume}{B530}},
  \bibinfo{pages}{167} (\bibinfo{year}{2002}), \eprint{hep-ph/0202074}\relax
\mciteBstWouldAddEndPuncttrue
\mciteSetBstMidEndSepPunct{\mcitedefaultmidpunct}
{\mcitedefaultendpunct}{\mcitedefaultseppunct}\relax
\EndOfBibitem
\bibitem[{\citenamefont{Xing}(2002)}]{Xing:2002sw}
\bibinfo{author}{\bibfnamefont{Z.-Z.} \bibnamefont{Xing}},
  \bibinfo{journal}{Phys. Lett.} \textbf{\bibinfo{volume}{B533}},
  \bibinfo{pages}{85} (\bibinfo{year}{2002}), \eprint{hep-ph/0204049}\relax
\mciteBstWouldAddEndPuncttrue
\mciteSetBstMidEndSepPunct{\mcitedefaultmidpunct}
{\mcitedefaultendpunct}{\mcitedefaultseppunct}\relax
\EndOfBibitem
\bibitem[{\citenamefont{Harrison and Scott}(2002)}]{Harrison:2002kp}
\bibinfo{author}{\bibfnamefont{P.~F.} \bibnamefont{Harrison}} \bibnamefont{and}
  \bibinfo{author}{\bibfnamefont{W.~G.} \bibnamefont{Scott}},
  \bibinfo{journal}{Phys. Lett.} \textbf{\bibinfo{volume}{B535}},
  \bibinfo{pages}{163} (\bibinfo{year}{2002}), \eprint{hep-ph/0203209}\relax
\mciteBstWouldAddEndPuncttrue
\mciteSetBstMidEndSepPunct{\mcitedefaultmidpunct}
{\mcitedefaultendpunct}{\mcitedefaultseppunct}\relax
\EndOfBibitem
\bibitem[{\citenamefont{Schwetz
  et~al.}(2011{\natexlab{a}})\citenamefont{Schwetz, Tortola, and
  Valle}}]{Schwetz:2011zk}
\bibinfo{author}{\bibfnamefont{T.}~\bibnamefont{Schwetz}},
  \bibinfo{author}{\bibfnamefont{M.}~\bibnamefont{Tortola}}, \bibnamefont{and}
  \bibinfo{author}{\bibfnamefont{J.~W.~F.} \bibnamefont{Valle}}
  (\bibinfo{year}{2011}{\natexlab{a}}), \eprint{1108.1376}\relax
\mciteBstWouldAddEndPuncttrue
\mciteSetBstMidEndSepPunct{\mcitedefaultmidpunct}
{\mcitedefaultendpunct}{\mcitedefaultseppunct}\relax
\EndOfBibitem
\bibitem[{\citenamefont{Schwetz
  et~al.}(2011{\natexlab{b}})\citenamefont{Schwetz, Tortola, and
  Valle}}]{Schwetz:2011qt}
\bibinfo{author}{\bibfnamefont{T.}~\bibnamefont{Schwetz}},
  \bibinfo{author}{\bibfnamefont{M.}~\bibnamefont{Tortola}}, \bibnamefont{and}
  \bibinfo{author}{\bibfnamefont{J.~W.~F.} \bibnamefont{Valle}},
  \bibinfo{journal}{New J. Phys.} \textbf{\bibinfo{volume}{13}},
  \bibinfo{pages}{063004} (\bibinfo{year}{2011}{\natexlab{b}}),
  \eprint{1103.0734}\relax
\mciteBstWouldAddEndPuncttrue
\mciteSetBstMidEndSepPunct{\mcitedefaultmidpunct}
{\mcitedefaultendpunct}{\mcitedefaultseppunct}\relax
\EndOfBibitem
\bibitem[{\citenamefont{Hettmansperger
  et~al.}(2011)\citenamefont{Hettmansperger, Lindner, and
  Rodejohann}}]{Hettmansperger:2011bt}
\bibinfo{author}{\bibfnamefont{H.}~\bibnamefont{Hettmansperger}},
  \bibinfo{author}{\bibfnamefont{M.}~\bibnamefont{Lindner}}, \bibnamefont{and}
  \bibinfo{author}{\bibfnamefont{W.}~\bibnamefont{Rodejohann}},
  \bibinfo{journal}{JHEP} \textbf{\bibinfo{volume}{1104}}, \bibinfo{pages}{123}
  (\bibinfo{year}{2011}), \eprint{1102.3432}\relax
\mciteBstWouldAddEndPuncttrue
\mciteSetBstMidEndSepPunct{\mcitedefaultmidpunct}
{\mcitedefaultendpunct}{\mcitedefaultseppunct}\relax
\EndOfBibitem
\bibitem[{\citenamefont{Rodejohann}(2011)}]{Rodejohann:2011mu}
\bibinfo{author}{\bibfnamefont{W.}~\bibnamefont{Rodejohann}},
  \bibinfo{journal}{Int. J. Mod. Phys.} \textbf{\bibinfo{volume}{E20}},
  \bibinfo{pages}{1833} (\bibinfo{year}{2011}), \eprint{1106.1334}\relax
\mciteBstWouldAddEndPuncttrue
\mciteSetBstMidEndSepPunct{\mcitedefaultmidpunct}
{\mcitedefaultendpunct}{\mcitedefaultseppunct}\relax
\EndOfBibitem
\bibitem[{\citenamefont{Boyarsky et~al.}(2009)\citenamefont{Boyarsky,
  Ruchayskiy, and Shaposhnikov}}]{Boyarsky:2009ix}
\bibinfo{author}{\bibfnamefont{A.}~\bibnamefont{Boyarsky}},
  \bibinfo{author}{\bibfnamefont{O.}~\bibnamefont{Ruchayskiy}},
  \bibnamefont{and}
  \bibinfo{author}{\bibfnamefont{M.}~\bibnamefont{Shaposhnikov}},
  \bibinfo{journal}{Ann.Rev.Nucl.Part.Sci.} \textbf{\bibinfo{volume}{59}},
  \bibinfo{pages}{191} (\bibinfo{year}{2009}), \eprint{0901.0011}\relax
\mciteBstWouldAddEndPuncttrue
\mciteSetBstMidEndSepPunct{\mcitedefaultmidpunct}
{\mcitedefaultendpunct}{\mcitedefaultseppunct}\relax
\EndOfBibitem
\bibitem[{\citenamefont{Abe et~al.}(2012)}]{Abe:2011fz}
\bibinfo{author}{\bibfnamefont{Y.}~\bibnamefont{Abe}} \bibnamefont{et~al.}
  (\bibinfo{collaboration}{DOUBLE-CHOOZ Collaboration}),
  \bibinfo{journal}{Phys.Rev.Lett.} \textbf{\bibinfo{volume}{108}},
  \bibinfo{pages}{131801} (\bibinfo{year}{2012}), \eprint{1112.6353}\relax
\mciteBstWouldAddEndPuncttrue
\mciteSetBstMidEndSepPunct{\mcitedefaultmidpunct}
{\mcitedefaultendpunct}{\mcitedefaultseppunct}\relax
\EndOfBibitem
\bibitem[{\citenamefont{An et~al.}(2012)}]{An:2012eh}
\bibinfo{author}{\bibfnamefont{F.}~\bibnamefont{An}} \bibnamefont{et~al.}
  (\bibinfo{collaboration}{DAYA-BAY Collaboration}),
  \bibinfo{journal}{Phys.Rev.Lett.} \textbf{\bibinfo{volume}{108}},
  \bibinfo{pages}{171803} (\bibinfo{year}{2012}), \eprint{1203.1669}\relax
\mciteBstWouldAddEndPuncttrue
\mciteSetBstMidEndSepPunct{\mcitedefaultmidpunct}
{\mcitedefaultendpunct}{\mcitedefaultseppunct}\relax
\EndOfBibitem
\bibitem[{\citenamefont{Ahn et~al.}(2012)}]{Ahn:2012nd}
\bibinfo{author}{\bibfnamefont{J.}~\bibnamefont{Ahn}} \bibnamefont{et~al.}
  (\bibinfo{collaboration}{RENO collaboration}),
  \bibinfo{journal}{Phys.Rev.Lett.} \textbf{\bibinfo{volume}{108}},
  \bibinfo{pages}{191802} (\bibinfo{year}{2012}), \eprint{1204.0626}\relax
\mciteBstWouldAddEndPuncttrue
\mciteSetBstMidEndSepPunct{\mcitedefaultmidpunct}
{\mcitedefaultendpunct}{\mcitedefaultseppunct}\relax
\EndOfBibitem
\bibitem[{\citenamefont{Fogli et~al.}(2012)\citenamefont{Fogli, Lisi, Marrone,
  Montanino, Palazzo et~al.}}]{Fogli:2012ua}
\bibinfo{author}{\bibfnamefont{G.}~\bibnamefont{Fogli}},
  \bibinfo{author}{\bibfnamefont{E.}~\bibnamefont{Lisi}},
  \bibinfo{author}{\bibfnamefont{A.}~\bibnamefont{Marrone}},
  \bibinfo{author}{\bibfnamefont{D.}~\bibnamefont{Montanino}},
  \bibinfo{author}{\bibfnamefont{A.}~\bibnamefont{Palazzo}},
  \bibnamefont{et~al.} (\bibinfo{year}{2012}), \eprint{1205.5254}\relax
\mciteBstWouldAddEndPuncttrue
\mciteSetBstMidEndSepPunct{\mcitedefaultmidpunct}
{\mcitedefaultendpunct}{\mcitedefaultseppunct}\relax
\EndOfBibitem
\bibitem[{\citenamefont{Altarelli and Feruglio}(2005)}]{Altarelli:2005yp}
\bibinfo{author}{\bibfnamefont{G.}~\bibnamefont{Altarelli}} \bibnamefont{and}
  \bibinfo{author}{\bibfnamefont{F.}~\bibnamefont{Feruglio}},
  \bibinfo{journal}{Nucl. Phys.} \textbf{\bibinfo{volume}{B720}},
  \bibinfo{pages}{64} (\bibinfo{year}{2005}), \eprint{hep-ph/0504165}\relax
\mciteBstWouldAddEndPuncttrue
\mciteSetBstMidEndSepPunct{\mcitedefaultmidpunct}
{\mcitedefaultendpunct}{\mcitedefaultseppunct}\relax
\EndOfBibitem
\end{mcitethebibliography}
\bibliographystyle{apsrevM}

\end{document}